\documentclass[twocolumn]{aastex701}

\usepackage{amsmath,amssymb}
\usepackage{amsthm}
\usepackage{gensymb}

\begin{document}

\title{Solitary Alfv\'en Waves}

\author[0000-0001-9570-5975]{Zesen Huang}
\affiliation{Department of Earth, Planetary, and Space Sciences, University of California, Los Angeles}
\email{zesenhuang@g.ucla.edu}

\author[0000-0002-2381-3106]{Marco Velli}
\affiliation{Department of Earth, Planetary, and Space Sciences, University of California, Los Angeles}
\email{mvelli@ucla.edu}

\author[0000-0002-2582-7085]{Chen Shi}
\affiliation{Department of Physics, Auburn University}
\email{chs0090@auburn.edu}

\author[0009-0009-0162-2067]{Yuliang Ding}
\affiliation{Department of Earth, Planetary, and Space Sciences, University of California, Los Angeles}
\email{dingyl@g.ucla.edu}

\begin{abstract}
We present the solitary Alfv\'en wave as an ideal nonlinear Alfv\'enic solution in the solitary far-field limit and construct a three-dimensional numerical model---an \emph{Alfv\'enon}. The model is characterized by an unperturbed far field, quasi-constant $|\boldsymbol{B}|$, and open field-line topology. Direct MHD simulations of the Alfv\'enon show coherent finite-time propagation, confirming that it behaves as a nonlinear solitary Alfv\'enic solution under ideal MHD evolution.
\end{abstract}


\section{Introduction}\label{sec:intro}

Since the introduction of Alfv\'en waves in 1942~\citep{alfven_existence_1942}, these fundamental magnetohydrodynamic (MHD) oscillations have been invoked to explain a wide range of phenomena in astrophysical~\citep{mckee_alfven_1995}, space~\citep{coleman_turbulence_1968,unti_alfven_1968,belcher_large-amplitude_1971}, and laboratory plasmas~\citep{gekelman_laboratory_1997}. Early solar wind observations showed that Alfv\'en waves are large-amplitude fluctuations characterized by quasi-constant $|\boldsymbol{B}|$, $p$, and $\rho$ \citep{belcher_large-amplitude_1971}. Recent observations from the Parker Solar Probe (PSP)~\citep{fox_solar_2016,raouafi_parker_2023} reveal that the pristine solar wind in the upper corona~\citep{kasper_parker_2021} is permeated by large-amplitude Alfv\'enic fluctuations. Intriguingly, these fluctuations manifest as solitary, large-amplitude magnetic field reversals with quasi-constant $|\boldsymbol{B}|$, accompanied by one-sided anti-sunward proton jets that exhibit near-perfect Alfv\'enic correlation~\citep{bale_highly_2019}. They also display field-aligned electron strahls characterized by one-sided pitch angles both inside and outside the field reversal regions, suggesting topologically open magnetic field lines~\citep{kasper_alfvenic_2019}. Consequently, some of these fluctuations are termed magnetic ``switchbacks'' (see also the recent review by \citet{badman_properties_2026}).

Due to the constancy of $|\boldsymbol{B}|$, these Alfv\'enic fluctuations are termed spherically polarized Alfv\'en waves (SPAWs)---exact nonlinear solutions of the ideal MHD equations~\citep{goldstein_theory_1974,hollweg_transverse_1974}. Conventionally, the background magnetic field $\boldsymbol{B}_0$ in SPAWs is defined as the ensemble average of $\boldsymbol{B}$ (and hence by construction $|\boldsymbol{B}_0| < |\boldsymbol{B}|$). This definition, however, is problematic: (1) In \textit{in situ} observations, the Alfv\'enic correlation $\delta \boldsymbol{B}/\sqrt{\mu_0 \rho}=\pm \delta \boldsymbol{u}$ uniquely determines the perturbative part of $\boldsymbol{B}$. Hence the local $\boldsymbol{B}_0$ should be obtained as $\boldsymbol{B}-\delta \boldsymbol{B}$. (2) The ensemble average of $\boldsymbol{B}$ is inherently arbitrary, as it can vary substantially within a single solar wind stream, and consequently so does the Alfv\'en velocity $\boldsymbol{V}_A = \boldsymbol{B}_0/\sqrt{\mu_0 \rho}$. \citet{gosling_one-sided_2009} demonstrated that Alfv\'en waves in the solar wind are predominantly one-sided. This concept was subsequently extended by \citet{matteini_dependence_2014} to explain local radial velocity enhancements in Ulysses observations, where $\boldsymbol{B}_0$ is defined on the constant sphere of $|\boldsymbol{B}|$. The groundbreaking PSP observations in the upper corona have since reinforced this perspective: Alfv\'enic fluctuations in the most pristine solar wind appear as \emph{solitary} perturbations on top of an otherwise \emph{unperturbed} coronal magnetic field. By removing the perturbative component of $\boldsymbol{B}$ constructed from the proton flow $\boldsymbol{u}$ via the Alfv\'enic correlation, the background $\boldsymbol{B}_0$ emerges naturally as the unperturbed coronal field. Consequently, Alfv\'en waves should be derivable from the ideal MHD equations as \textit{solitary} solutions assuming only incompressibility (quasi-constant $|\boldsymbol{B}|$, $\rho$, and $p$) without \textit{a priori} assumptions about $\boldsymbol{B}_0$.

However, modeling SPAWs presents substantial challenges. Constructing a SPAW model involves three steps: (1) finding a magnetic field $\boldsymbol{B}$ satisfying both $\nabla \cdot \boldsymbol{B}=0$ and $|\boldsymbol{B}|\simeq \mathrm{const}$; (2) decomposing $\boldsymbol{B}$ into a background component $\boldsymbol{B}_0$ and a fluctuating component $\boldsymbol{B}_1$; and (3) constructing the velocity perturbation $\boldsymbol{u}_1$ from $\boldsymbol{B}_1$ via the Alfv\'enic correlation. Step (1) alone poses a highly non-trivial mathematical problem. \citet{barnes_nonexistence_1976} demonstrated the non-existence of 2D solutions (with $\boldsymbol{B}$ restricted to a plane). Furthermore, assuming a constant far field (i.e., solitary solutions), 2.5D configurations (two spatial coordinates, three vector components) exist only when the vector field contains topologically closed regions (see Appendix~\ref{appendix:A}). Therefore, maintaining open field-line topology requires $\boldsymbol{B}$ to be genuinely 3D. Steps (2--3) are also non-trivial because $\boldsymbol{u}_1$ depends on $\boldsymbol{B}_1$, which in turn depends on the choice of $\boldsymbol{B}_0$. This decomposition is well-defined only for solitary solutions; otherwise, the determination of $\boldsymbol{B}_0$ remains ambiguous. Several studies have attempted to construct SPAWs. \citet{tenerani_magnetic_2020} and \citet{shi_analytic_2024} developed analytic solitary 2.5D switchback models, which necessarily contain topologically closed regions. \citet{squire_construction_2022} and \citet{shoda_turbulent_2021} constructed 3D turbulent switchbacks that lack spatial isolation, precluding a well-defined separation between $\boldsymbol{B}_0$ and $\boldsymbol{B}_1$.

In this study, we derive solitary Alfv\'en waves from the ideal MHD equations assuming only incompressibility (quasi-constant $|\boldsymbol{B}|$, $\rho$, and $p$), with $\boldsymbol{B}_0$ naturally emerging as the constant far field. Based on this solution, we construct a solitary Alfv\'en wave model---an \emph{Alfv\'enon}---characterized by quasi-constant $|B|$ and open field-line topology. The remainder of this paper is organized as follows: Section~\ref{sec:equations} presents the solitary Alfv\'en wave solution. Section~\ref{sec:model} constructs the Alfv\'enon. Section~\ref{sec:simulation} presents results from MHD simulations of the Alfv\'enon. Finally, Section~\ref{sec:discussion} discusses and concludes the study.

\section{Solitary Alfv\'en Waves}\label{sec:equations}
We begin from the ideal MHD equations assuming adiabatic processes:
\begin{align}
    \frac{\partial \rho}{\partial t} + \nabla\cdot (\rho \boldsymbol u) &= 0,\label{eq:mhd1}\\
    \rho\left[\frac{\partial \boldsymbol u}{\partial t} + (\boldsymbol u \cdot \nabla) \boldsymbol u \right] &= -\nabla \left(p+\frac{B^2}{2\mu_0}\right) + \frac{1}{\mu_0} (\boldsymbol B\cdot \nabla) \boldsymbol B,\label{eq:mhd2}\\
    \frac{\partial \boldsymbol B}{\partial t} &= \nabla \times (\boldsymbol u \times \boldsymbol B),\label{eq:mhd3}\\
    p \rho^{-\gamma} &= \mathrm{const},\label{eq:mhd4}\\
    \nabla \cdot \boldsymbol B &= 0,\label{eq:mhd5}
\end{align}
where $\rho$ is the plasma density, $\boldsymbol u$ is the flow velocity, $p$ is the pressure, $\boldsymbol B$ is the magnetic field, $\gamma$ is the adiabatic index, and $\mu_0$ is the vacuum permeability. To proceed, we assume incompressibility: constant $|\boldsymbol B|$, $\rho$ and $p$. Under these assumptions, we look for \emph{solitary Alfv\'enic} solution of $\boldsymbol B(\boldsymbol r, t)$ and $\boldsymbol u(\boldsymbol r, t)$. 

\begin{figure}[t]
    \centering
    \includegraphics[width=0.48\columnwidth]{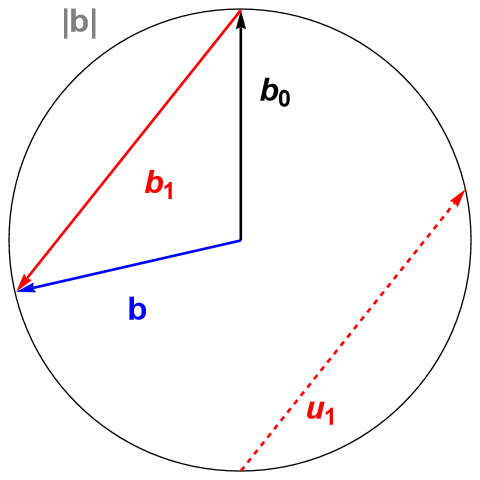}
    \hspace{0.0\columnwidth}
    \includegraphics[width=0.48\columnwidth]{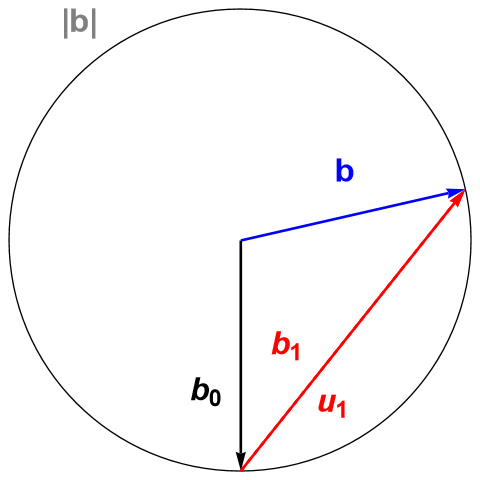}
    \caption{Constant $|B|$ constraint for forward propagating SPAWs. \emph{Left}: forward $\boldsymbol b_0$. \emph{Right}: backward $\boldsymbol b_0$.}
    \label{fig:model}
\end{figure}

First, we make the conversion: $\boldsymbol b = \boldsymbol B/\sqrt{\mu_0 \rho}$. Incompressibility guarantees $\nabla\cdot \boldsymbol u = 0$ and $\nabla \cdot \boldsymbol b = 0$. Consequently, Eqs.~\eqref{eq:mhd2}--\eqref{eq:mhd3} reduce to:
\begin{align}
    % \nabla\cdot \boldsymbol u &= 0, \label{eq:simplified1}\\
    % \nabla\cdot \boldsymbol b &= 0, \label{eq:simplified2}\\
    \frac{\partial \boldsymbol u}{\partial t}  &=  (\boldsymbol b\cdot \nabla) \boldsymbol b - (\boldsymbol u \cdot \nabla) \boldsymbol u,\label{eq:simplified3}\\
    \frac{\partial \boldsymbol b}{\partial t} &= (\boldsymbol b \cdot \nabla) \boldsymbol u - (\boldsymbol u \cdot \nabla) \boldsymbol b.\label{eq:simplified4}
\end{align}
Under the constant-$|\boldsymbol b|$ constraint, separating $\boldsymbol b$ into a DC component $\boldsymbol b_0$ and an AC perturbation $\boldsymbol b_1$ is nontrivial. Previous derivations of shear, circularly polarized, and large-amplitude/spherically polarized Alfv\'en waves~\citep{alfven_existence_1942,walen_theory_1944,goldstein_theory_1974,hollweg_transverse_1974} either leave the unperturbed state unspecified or set $\boldsymbol B_0=\langle\boldsymbol B\rangle$; such an averaged background depends on the averaging window and generally does not satisfy $|\boldsymbol B_0|=|\boldsymbol B|$, so it provides neither a deterministic AC/DC separation nor a unique Alfv\'en velocity. For a solitary wave packet, the separation is instead fixed by the far-field condition
\begin{equation}
    \boldsymbol b_1\to 0, \qquad \boldsymbol b_0 \equiv \lim_{r\to\infty}\boldsymbol b(\boldsymbol r,t).
\end{equation}
Thus $\boldsymbol b_0$ is the unique unperturbed (ground) state. Since $|\boldsymbol b|$ is constant in the ideal formulation, $|\boldsymbol b_0|=|\boldsymbol b|$, and $\boldsymbol b_0$ uniquely defines the Alfv\'en velocity; $\boldsymbol b_1=\boldsymbol b-\boldsymbol b_0$ is defined with respect to this state. Geometrically, the tip of $\boldsymbol b$ remains on the constant-$|\boldsymbol b|$ sphere, so the perturbation $\boldsymbol b_1$ is constrained to the corresponding sphere defined by $\boldsymbol b_0$, as illustrated in Fig.~\ref{fig:model}. Similarly, we decompose $\boldsymbol u$ into $\boldsymbol u_0 + \boldsymbol u_1$. \emph{Alfv\'enic} solution dictates: $\boldsymbol u_1 = \pm \boldsymbol b_1$.  Because $\boldsymbol b$ is Galilean invariant, without loss of generality, we can transform into a frame where $\boldsymbol u_0 = 0$. In this frame, Eqs.~\eqref{eq:simplified3}--\eqref{eq:simplified4} reduce to:
\begin{align}
    \frac{\partial \boldsymbol u_1}{\partial t}  &=  \boldsymbol b_0 \cdot \nabla \boldsymbol b_1, \label{eq:alfven1}\\
    \frac{\partial \boldsymbol b_1}{\partial t} &= \boldsymbol b_0 \cdot \nabla\boldsymbol u_1, \label{eq:alfven2}
\end{align}
where $\boldsymbol b_0 = \boldsymbol B_0/\sqrt{\mu_0 \rho}$ represents the Alfv\'en velocity. These yield the wave equations of solitary Alfv\'en waves:
\begin{align}
    \frac{\partial^2 \boldsymbol u_1}{\partial t^2}  &=  (\boldsymbol b_0 \cdot \nabla)^2 \boldsymbol u_1, \label{eq:wave1}\\
    \frac{\partial^2 \boldsymbol b_1}{\partial t^2}  &=  (\boldsymbol b_0 \cdot \nabla)^2 \boldsymbol b_1. \label{eq:wave2}
\end{align}

This derivation should be read as the ideal constant-$|\boldsymbol B|$ limit. In a finite domain, a finite-amplitude localized structure that reduces or reverses $B_x$ cannot remain both strictly solitary and perfectly constant in $|\boldsymbol B|$ while satisfying exact solenoidality and magnetic-flux conservation. The missing axial flux inside the perturbed region must be compensated by a slight compression or enhancement of neighboring field lines, so the numerical Alfv\'enon constructed below is a controlled quasi-constant-$|\boldsymbol B|$ realization of the ideal solitary Alfv\'en-wave solution. The detailed flux-balance argument is given in Section~\ref{subsec:subtleties}.

An important consequence follows from Eqs.~\eqref{eq:alfven1}--\eqref{eq:alfven2}. Assuming $\boldsymbol b_0 = b_0 \hat{x}$ and $\boldsymbol b_1 = -\boldsymbol u_1$, Eq.~\eqref{eq:alfven1} becomes
\begin{align}
    \left(\frac{\partial}{\partial t}+b_0 \frac{\partial }{\partial x}\right)\boldsymbol b_1 = 0, \label{eq:propagation}
\end{align}
describing a forward-propagating ($+\hat{x}$) wave $\boldsymbol b_1(x - b_0 t)$. Similarly, when $\boldsymbol b_0 = -b_0 \hat{x}$, forward propagation requires $\boldsymbol b_1 = \boldsymbol u_1$. Thus for forward-propagating waves, $u_{1x}$ is always positive irrespective of the sign of $\boldsymbol b_0$ (Fig.~\ref{fig:model}). This explains the one-sided anti-sunward proton jets associated with the SPAWs/switchbacks in the solar wind~\citep{gosling_one-sided_2009,matteini_dependence_2014,bale_highly_2019,kasper_alfvenic_2019,badman_properties_2026}.

\section{The Alfv\'enon Model}\label{sec:model}

Despite the simplicity of the solitary Alfv\'en waves, constructing an Alfv\'enon is non-trivial. Without loss of generality, we seek a magnetic field $\boldsymbol{B}(x,y,z)$ satisfying four constraints: (1) $|\boldsymbol{B}|\simeq 1$, (2) $\nabla\cdot \boldsymbol{B}=0$, (3) open field-line topology, and (4) constant far field (solitary). Constraints (1) and (4) are \emph{soft} (approximate), while (2) and (3) are \emph{hard} (exact).

\subsection{The Iterative Algorithm}
Given an arbitrary three-dimensional vector field $\boldsymbol{F}(x,y,z)$, we apply the Helmholtz-Hodge decomposition:
\begin{align}
\boldsymbol F = \nabla \varphi + \nabla \times \boldsymbol A,
\end{align}
where $\varphi$ is solved from the Poisson equation $\nabla^2 \varphi = \nabla \cdot \boldsymbol F$. The divergence can be removed via:
\begin{align}
\boldsymbol G = \boldsymbol F - \nabla \varphi. \label{eq:remove_divergence}
\end{align}
The resulting field $\boldsymbol G$ is then normalized to a unit vector field:
\begin{align}
    \boldsymbol F' = \frac{\boldsymbol G}{|\boldsymbol G|}. \label{eq:normalization}
\end{align}
Such normalization reintroduces nonzero divergence. The iteration can therefore be viewed geometrically as an alternating-projection-type feasibility problem between two constraint sets, related to projection algorithms for convex feasibility and to alternating-projection methods for both convex and non-convex settings \citep{bauschke_projection_1996,escalante_alternating_2011,lewis_local_2009}. The Helmholtz-Hodge step is the $L^2$-orthogonal projection onto the solenoidal subspace within the chosen periodic Fourier representation and fixed mean field, whereas the normalization step is a pointwise projection onto the unit-$|\boldsymbol B|$ manifold, i.e., a product of spheres wherever $|\boldsymbol G|$ is nonzero. Because the unit-magnitude constraint set is non-convex, standard convergence theorems for alternating projections onto convex sets do not provide a global guarantee for the present algorithm. We therefore do not claim convergence for arbitrary initial fields; instead, convergence is treated as an empirical property of the smooth localized seeds considered here. The basin of attraction, possible failure modes, and rigorous convergence criteria remain open mathematical questions. However, with appropriate initial conditions $\boldsymbol F$, alternating application of Eqs.~\eqref{eq:remove_divergence} and~\eqref{eq:normalization} empirically converges. Denoting the fields at the $n$-th iteration by $\boldsymbol F_n$ and $\boldsymbol G_n$, we have:

\begin{align}
    \boldsymbol G_{n} &= \boldsymbol F_{n} - \nabla \varphi_n,  \label{eq:F2G}\\
    \boldsymbol F_{n+1} &= \frac{\boldsymbol G_n}{|\boldsymbol G_n|} \label{eq:G2F},
\end{align}
where $\nabla^2 \varphi_n = \nabla \cdot \boldsymbol F_n$. Equations~\eqref{eq:F2G} and \eqref{eq:G2F} together constitute the iterative algorithm, hereafter referred to simply as the algorithm. In practice, the Helmholtz-Hodge decomposition is implemented in Fourier space (and thus assuming periodic boundary conditions) by solving the Poisson equation:
\begin{align}
    \widehat{\nabla\cdot \boldsymbol F}=i \boldsymbol k \cdot \widehat{\boldsymbol{F}}=\widehat{\Delta\varphi}=-k^2\widehat{\varphi}.
\end{align}
and thus the potential field is obtained to be:
\begin{align}
    \widehat{\varphi} = -\frac{i \boldsymbol k \cdot \widehat{\boldsymbol{F}}}{k^2},
\end{align}
where the zero-$k$ singular point $\widehat{\varphi}(0)$ is set to be an arbitrary value. The solenoidal field can then be obtained:
\begin{align}
    \begin{aligned}
    \widehat{\boldsymbol{F_\perp}}=\widehat{\boldsymbol{F}}-\widehat{\nabla \varphi}&=\widehat{\boldsymbol{F}}-\left[i\boldsymbol k \left(-\frac{i \boldsymbol k \cdot \widehat{\boldsymbol{F}}}{k^2}\right)\right]\\
    &= \widehat{\boldsymbol{F}}-\boldsymbol k \left(\frac{\boldsymbol k \cdot \widehat{\boldsymbol{F}}}{k^2}\right).
    \end{aligned}
\end{align}
$\widehat{\boldsymbol{F_\perp}}$ can then be inverse-transformed into real space, which is denoted as $\boldsymbol G$. The values of $\widehat{\boldsymbol{F}}$ and $\widehat{\boldsymbol{F_\perp}}$ on the Nyquist planes are set to zero to avoid aliasing. Otherwise, significant divergence persists in $\boldsymbol G$ for $|k| \geq k_{\text{Nyquist}}$.

\subsection{Magnetic Field}

\begin{figure}[t]
    \centering
    \includegraphics[width =1.0\columnwidth]{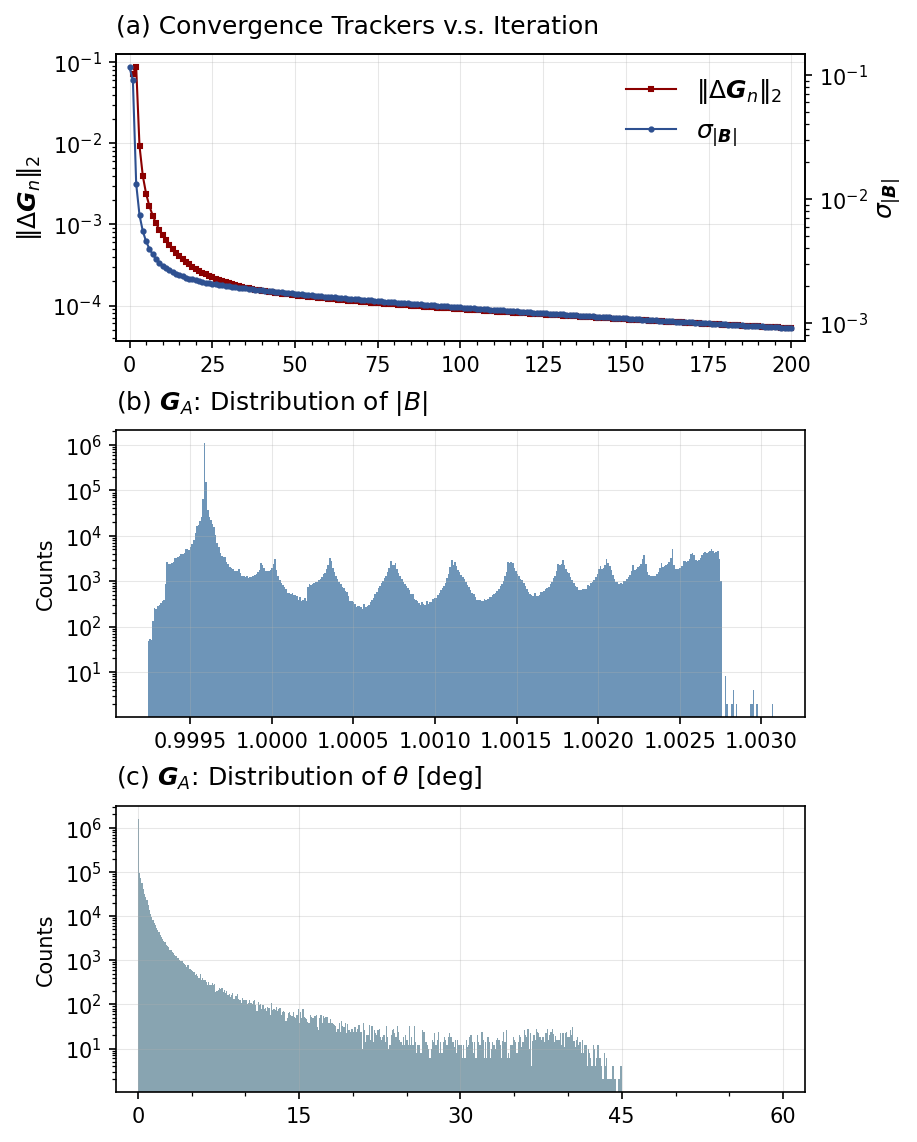}
    \caption{(a) $\|\Delta \boldsymbol{G}_n\|_2$ (left) and $\sigma_{|\boldsymbol{B}|}$ of $\boldsymbol{G}_n$ (right) versus iteration number. (b) Histogram of $|\boldsymbol{B}|$ of $\boldsymbol{G}_{A}$. (c) Histogram of $\theta$ of $\boldsymbol{G}_{A}$.}
    \label{fig:convergence}
\end{figure}

\begin{figure}[t]
    \centering
    \includegraphics[width = 1.0\columnwidth]{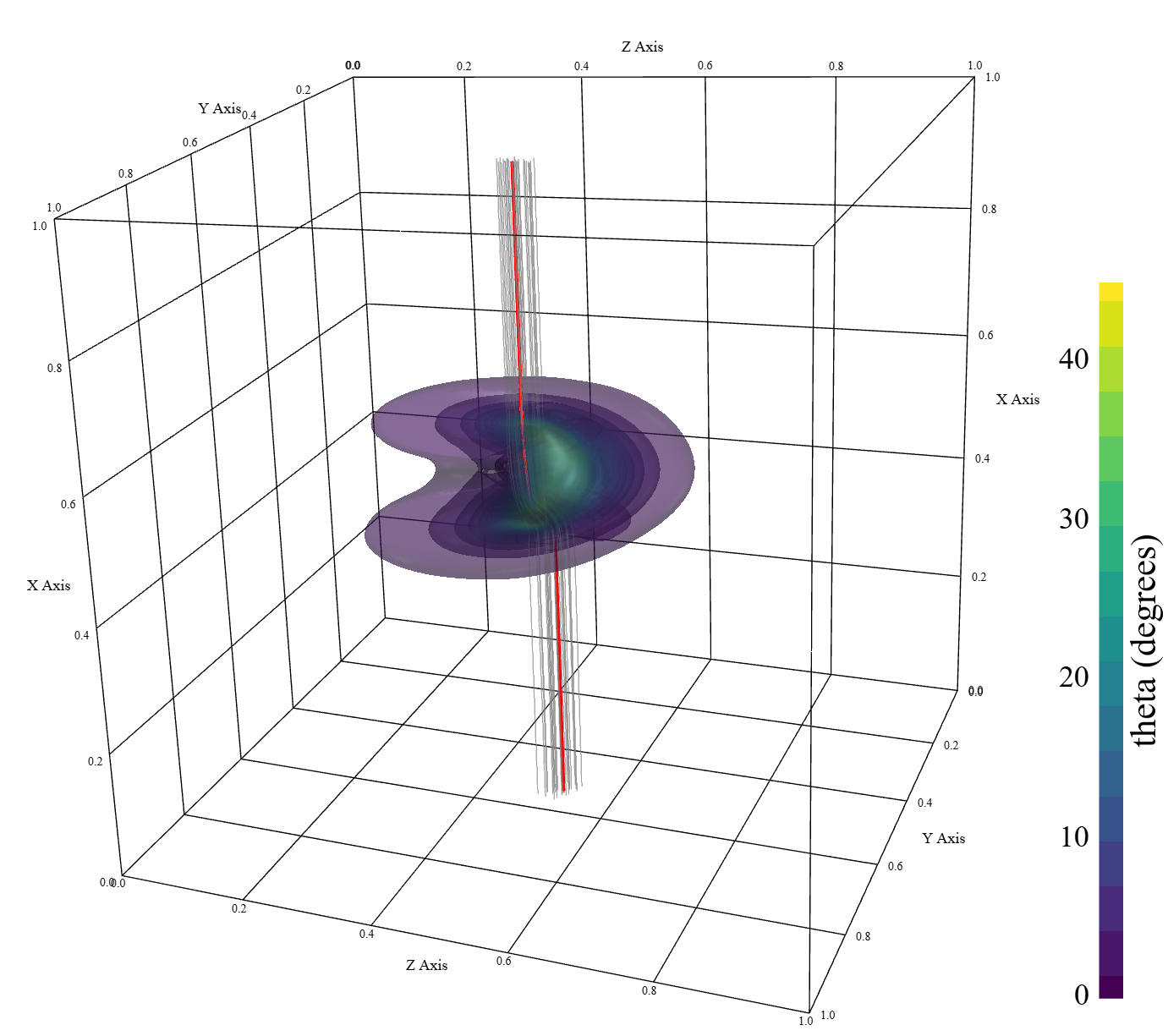}
    \caption{Contour of $\theta(x,y,z)$ of $\boldsymbol{G}_A$.}
    \label{fig:contour}
\end{figure}

We construct the Alfv\'enon model on a $128^3$ grid, with all spatial coordinates $x$, $y$ and $z$ ranging from 0 to 1. We start from an initial field:
\begin{align}
    \boldsymbol{F}_0 = \boldsymbol{B}_0 + A \cdot \left[
        \cos(\phi) \hat{y} + \sin(\phi) \hat{z}
    \right] \cdot \exp\left[-
        \frac{\Delta r^2}{2\sigma^2}
    \right]
\end{align}
where $\boldsymbol{B}_0 = (1,0,0)$, $A=10.0$, $\phi = 2\pi k_x x$, $k_x = 4$, $\Delta r = \sqrt{(x-0.5)^2+(y-0.5)^2+(z-0.5)^2}$, and $\sigma = 1/30$. Because of the Gaussian envelope, $\boldsymbol{F}_0$ is neither divergence-free nor of constant magnitude, making it a suitable input for the algorithm. Convergence is tracked with the vector field difference:

\begin{equation}
\|\Delta \boldsymbol{G}_n\|_2
= \sqrt{
\sum_{i,j,k}
\big|\boldsymbol{G}_n
      - \boldsymbol{G}_{n-1}\big|^2
\,\Delta x\,\Delta y\,\Delta z
},
\end{equation}
where $\Delta x = \Delta y = \Delta z = 1/128$. The results are shown in Figure~\ref{fig:convergence}(a). $\|\Delta \boldsymbol{G}_n\|_2$ drops rapidly within the first 25 iterations and then asymptotically converges. Similarly, the standard deviation $\sigma_{|\boldsymbol{B}|}$ of $\boldsymbol{G}_n$ decreases rapidly and becomes negligible as the iterations proceed. Due to this convergence, the iteration is stopped at $n=200$, and $\boldsymbol{G}_{200}$ is taken as a candidate magnetic field for the Alfv\'enon model, hereafter denoted as $\boldsymbol{G}_A$. We choose $\boldsymbol{G}_A$ as the candidate magnetic field because it is strictly solenoidal, satisfying the \emph{hard} constraint $\nabla\cdot\boldsymbol{B}=0$ up to machine precision ($10^{-10}$), while retaining a quasi-constant magnitude that satisfies the \emph{soft} constraint $|\boldsymbol{B}|\simeq 1$ to high accuracy.

The resulting field $\boldsymbol{G}_A$ exhibits a spatially localized, nontrivial twisting of otherwise unperturbed open magnetic field lines oriented along $+\hat{x}$. Outside the perturbed region, the field is approximately uniform, $\boldsymbol{B} \simeq \boldsymbol{B}_0 = (1,0,0)$. The distribution of $|\boldsymbol{B}|$ in $\boldsymbol{G}_A$ is shown in Fig.~\ref{fig:convergence}(b), demonstrating that $|\boldsymbol{B}|$ remains nearly constant on the unit sphere. To characterize the structure of the solution, we define the local deflection angle from $\boldsymbol{B}_0$ as
\[
\theta(x,y,z) = \cos^{-1}\!\left(\frac{B_x}{|\boldsymbol{B}|}\right).
\]
Contours of $\theta$ are displayed in Fig.~\ref{fig:contour}, with its distribution shown in Fig.~\ref{fig:convergence}(c). The maximum deflection angle, $\theta_{\max} \simeq 45.0^\circ$, occurs at the pair of reflection-symmetric grid points $P_1(ix = 59,\, iy = 70,\, iz = 64)$ and $P_2(ix = 69,\, iy = 58,\, iz = 64)$, where $ix$, $iy$, and $iz$ denote the integer indices on the $128^3$ grid (ranging from 0 to 127). The red field line shown in Fig.~\ref{fig:contour} passes through $P_1$, with neighboring field lines shown in gray. All field lines enter the computational domain through the $x = 0$ plane and exit through the $x = 1$ plane, confirming an open field-line topology throughout (see Appendix~\ref{appendix:B} for details of the field-line tracer). Outside the perturbed region, field lines remain essentially unperturbed. Field lines traversing the perturbed region undergo localized twisting while preserving field-line density ($|\boldsymbol{B}| \simeq 1$), and subsequently relax back to the unperturbed state $\boldsymbol{B}_0$ downstream. The $\theta_{\max} \simeq 45^\circ$ configuration studied here should be regarded as a moderate-amplitude prototype; the near-reversal, large-amplitude branch of solitary Alfv\'en waves will be treated separately in future work.

\subsection{The Alfv\'enon Model}

The quasi-constant far field $\boldsymbol{B}_0$ enables a clear separation of the perturbation $\boldsymbol{B}_1$ from $\boldsymbol{B}$. Based on the Alfv\'enic correlation, we construct $\boldsymbol{u}_1 = -\boldsymbol{b}_1 = -\boldsymbol{B}_1$ to ensure forward propagation along $\boldsymbol{B}_0$, where we have adopted normalized units with $\rho = 1$ and $\mu_0 = 1$, and set $\boldsymbol{u}_0 = 0$. These $\boldsymbol{B}$ and $\boldsymbol{u}$ fields serve as initial conditions for the MHD simulations.

\section{MHD Simulations}\label{sec:simulation}

\begin{figure*}[htp!]
    \centering
    \includegraphics[width = 1.0 \textwidth]{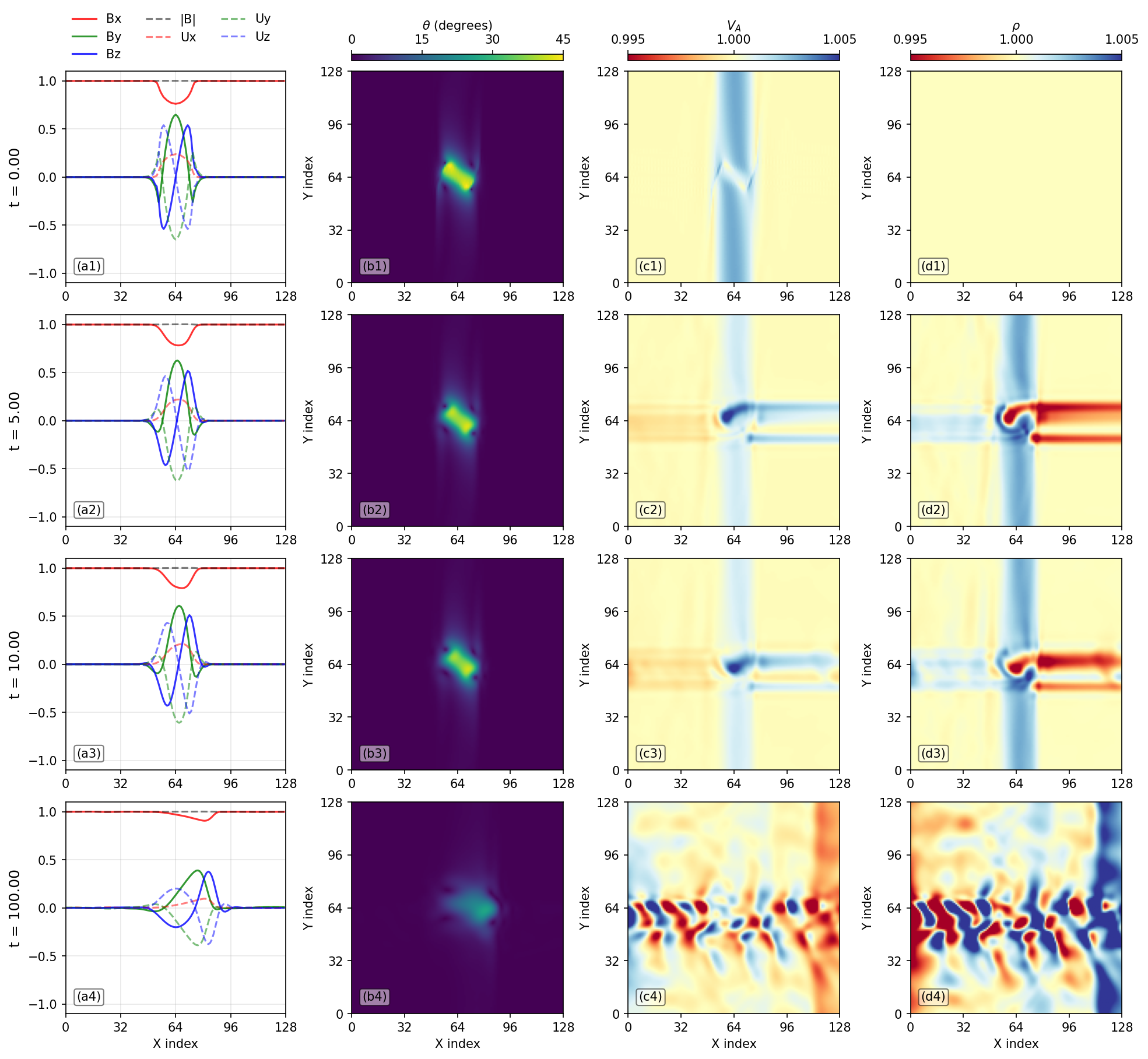}
    \caption{MHD simulation of the Alfv\'enon at $t=0.0, 5.0, 10.0, 100.0$. Column (a): One-dimensional profile at $iy=64$ and $iz = 64$. Column (b): Two-dimensional X-Y slice of $\theta$ at $iz = 64$. Column (c): $V_A = |\boldsymbol{B}|/\sqrt{\rho}$. Column (d): $\rho$. Rows 1--4 correspond to times $t=0.0$, 5.0, 10.0, and 100.0.}
    \label{fig:sim}
\end{figure*}

\begin{figure}[htp!]
    \centering
    \includegraphics[width=1.0\columnwidth]{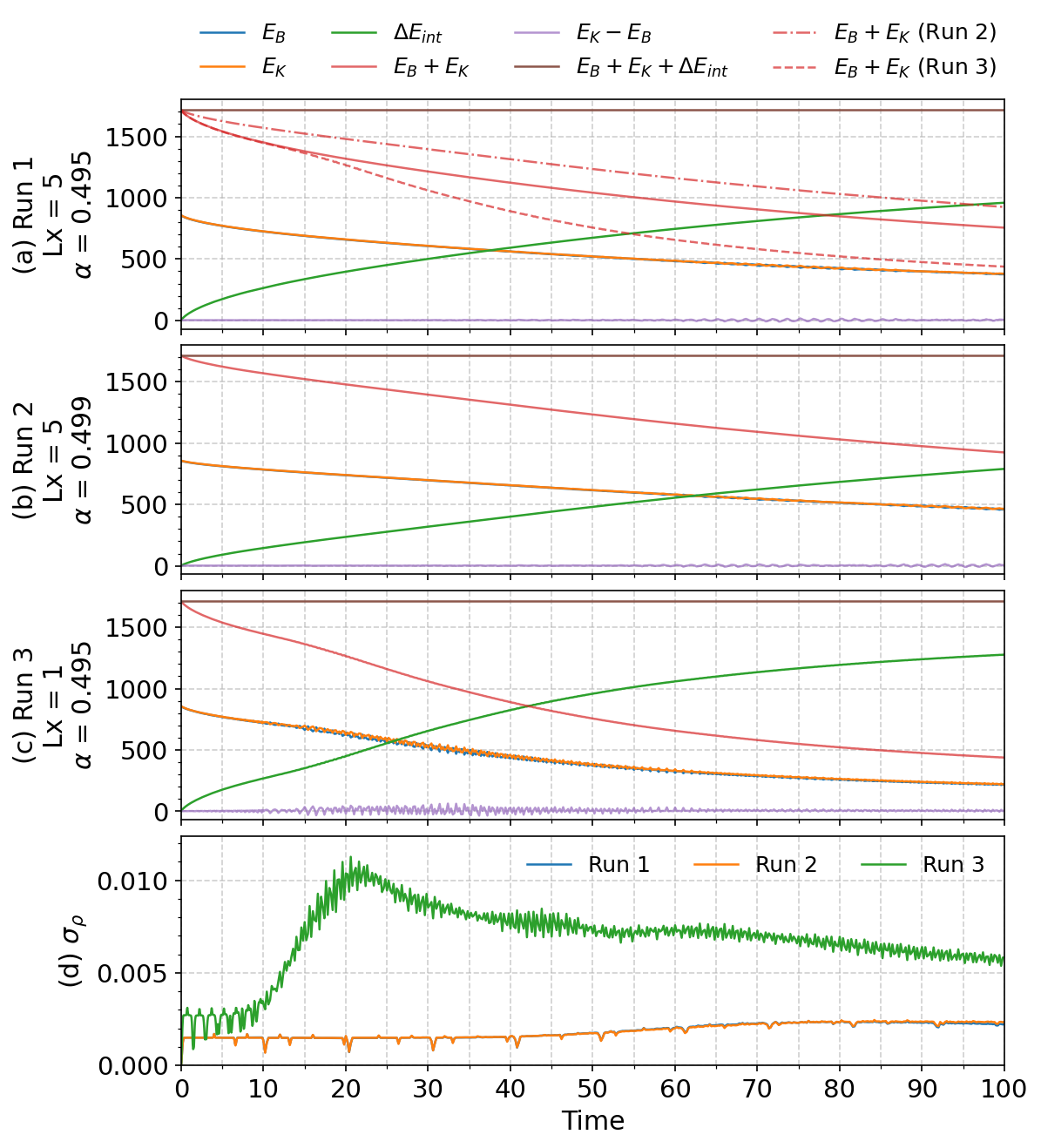}
    \caption{Comparison of three runs. (a) Run 1: $L_x=5$, $\alpha$ = 0.495. (b) Run 2: $L_x=5$, $\alpha$ = 0.499. (c) Run 3: $L_x=1$, $\alpha$ = 0.495. (d) Standard deviation of $\rho$.}
    \label{fig:energy}
\end{figure}

To validate that the Alfv\'enon model constitutes a numerical representation of solitary Alfv\'en waves, we perform MHD simulations using the LAPS code \citep{shi_laps_2024}, a pseudo-spectral solver for the ideal MHD equations with periodic boundary conditions. The plasma pressure is set to $p = 0.05$, corresponding to a plasma beta $\beta = 2p/B^2 = 0.1$ consistent with pristine solar wind conditions \citep{huang_dominance_2024}. A polytropic index $\gamma = 1.2$ is adopted to approximate realistic fast solar wind conditions \citep{shi_acceleration_2022}. Both viscosity and resistivity are set to zero; dissipation arises solely from numerical dealiasing. To minimize interactions between periodic images, the Alfv\'enon model (defined on a $128^3$ grid) is embedded within a larger domain by appending uniform $\boldsymbol{B}_0$ regions of size $128^3$ on either side along the $x$-direction. This produces an initial grid of $640\times128\times128$ spanning a domain of size $5\times 1\times 1$ ($L_x \times L_y \times L_z$), with the Alfv\'enon centered in the middle third.

Simulation results are presented in Figure~\ref{fig:sim} at times $t = 0.0$, $5.0$, $10.0$, and $100.0$, where $t$ denotes dimensionless simulation time. Column~(a) shows the one-dimensional wave-packet profile along the $x$-direction at $iy = 64$ and $iz = 64$. Profiles are shifted to a common reference position assuming a propagation speed of 1, with only the central 128 grid points displayed. All quantities are nondimensionalized such that $t = 1$ corresponds to the Alfv\'enon traversing one unit-length box, while crossing the full domain of length $L_x = 5$ requires $t = 5$. The near-perfect alignment of profiles demonstrates that the Alfv\'enon propagates at $V_A =|\boldsymbol{B}_0|/\sqrt{\rho} \simeq 1$ while maintaining spatial coherence with only mild relaxation. The perturbations preserve Alfv\'enic correlations in all three components, including $B_x$ and $u_x$ (along $\boldsymbol{B}_0$). Column~(b) displays the spatial distribution of $\theta$, revealing negligible nonlinear evolution up to $t = 10.0$ and gradual relaxation by $t = 100.0$.

The deformation of the Alfv\'enon arises primarily from phase mixing. Column~(c) displays the local $V_A$. Inside the Alfv\'enon, $V_A$ varies by approximately 1\%, producing differential phase speeds that gradually deform the wave packet. Furthermore, for slabs perpendicular to the $x$-axis, $V_A$ near the Alfv\'enon slightly exceeds that of the unperturbed region (by $\lesssim 0.5\%$), accounting for the gradual forward drift visible in column~(a). This $V_A$ distribution becomes progressively disrupted as the simulation proceeds, growing increasingly irregular by $t=100.0$ due to nonlinear evolution.

Nonlinear evolution also generates density fluctuations, violating perfect Alfv\'enicity. Column~(d) shows the density profiles. By construction, the initial Alfv\'enon contains no density perturbations. However, the model inevitably includes minor defects, such as high-frequency modes near the Nyquist frequency (suppressed by dealiasing procedures) and magnetic pressure imbalances arising from the slight non-constancy of $|\boldsymbol{B}|$. These defects induce density perturbations as the simulation progresses, which in turn drive nonlinear evolution. Notably, the spatial pattern of $\rho$ coincides with that of $V_A$, indicating that density variations dominate over $|\boldsymbol{B}|$ variations in causing the phase mixing.

To further investigate the sources of the observed relaxation, we performed two additional simulations beyond the initial run (Run 1: $L_x=5$, $\alpha=0.495$, where $\alpha$ is the controlling parameter of dealiasing option 2 in LAPS; dealiasing becomes less effective as $\alpha \to 0.5$; see Appendix~\ref{appendix:C} for details). These additional runs either suppress the dealiasing effect (Run 2: $L_x=5$, $\alpha=0.499$) or enhance the effects of periodic boundary conditions (Run 3: $L_x=1$, $\alpha=0.495$). Figure~\ref{fig:energy} presents the temporal evolution of energy diagnostics across all three runs: magnetic fluctuation energy $E_B = \frac{1}{2}|\boldsymbol{B}-\boldsymbol{B}_0|^2$; kinetic energy $E_K = \tfrac{1}{2} \rho |\boldsymbol{u}|^2$; incremental internal energy $\Delta E_{\mathrm{int}} = (p - p_0)/(\gamma - 1)$, where $p_0$ denotes the initial pressure; total fluctuation energy $E_B + E_K$; residual energy $E_r = E_K - E_B$, which vanishes for perfectly Alfv\'enic fluctuations; and total energy $E_B + E_K + \Delta E_{\mathrm{int}}$. All diagnostics are integrated over the simulation domain. For comparison, the total fluctuation energy from Runs 2 and 3 is also displayed in panel~(a).

Across all three runs, $E_B$ and $E_K$ gradually decrease while $\Delta E_{\mathrm{int}}$ increases, with total energy $E_B + E_K + \Delta E_{\mathrm{int}}$ conserved to high precision ($\sim 10^{-9}$). This confirms that fluctuation energy is converted entirely to internal energy via compressive work, with negligible numerical dissipation. Comparison of the runs reveals distinct influences from dealiasing and boundary effects. In Run 2, reduced dealiasing lowers the heating rate (dash-dotted line in Fig.~\ref{fig:energy}(a)), indicating that a substantial fraction of the compression is numerical. Nevertheless, perturbations in both Runs 1 and 2 remain perfectly Alfv\'enic (purple lines in panels a and b), with small density fluctuations (panel d). In Run 3, the shorter domain ($L_x = 1$) enhances nonlinear interactions from periodic boundaries (dashed line in panel a). While Runs 1 and 3 evolve nearly identically for the first 10 time units, Run 3 subsequently develops significant density fluctuations (panel d), generating non-zero residual energy (purple line in panel c) and accelerating heating (green line in panel c).

This domain-size dependence also clarifies the role of parametric decay instability (PDI) in the present simulations. PDI is naturally associated with extended, periodic, large-amplitude Alfv\'enic wave trains, whereas the object considered here is a localized solitary packet embedded in an otherwise unperturbed far field. In a periodic numerical domain, however, the packet is repeated by the boundary conditions; if the box is too short, these periodic images effectively increase the filling factor of the perturbation and make the system more wave-train-like, allowing PDI-like compressive sidebands to grow more efficiently. The comparison between the longer-domain runs and Run 3 shows that, in the finite-time periodic simulations presented here, the PDI-like signatures are strongly enhanced by finite-domain and periodic-image effects and are suppressed when the Alfv\'enon is more isolated from its periodic images. We therefore do not identify PDI as the dominant process controlling the simulated Alfv\'enon evolution over the present time interval. A full stability theory of isolated or open-boundary solitary Alfv\'en waves with respect to PDI is beyond the scope of this paper and will be addressed in future work.

Overall, these results demonstrate that the Alfv\'enon behaves as a nonlinear solitary Alfv\'enic solution under ideal MHD evolution.

\section{Discussion and Conclusions}\label{sec:discussion}

\subsection{Properties of Solitary Alfv\'en Waves}

In Section~\ref{sec:equations}, we derived solitary Alfv\'en waves from the ideal MHD equations under the assumption of incompressibility. Although Eqs.~\eqref{eq:wave1}--\eqref{eq:wave2} formally resemble those governing classical shear~\citep{alfven_existence_1942}, circularly polarized~\citep{walen_theory_1944}, or spherically polarized~\citep{goldstein_theory_1974} Alfv\'en waves, two critical distinctions arise:
\begin{itemize}
    \item $\boldsymbol{b}_0$ is the \emph{unperturbed} field, uniquely determined by the constant far field.
    \item $\boldsymbol{b}_1$ is \emph{solitary}, representing a localized perturbation of an otherwise unperturbed background field.
\end{itemize}

These distinctions lead to three non-trivial consequences: First, $\boldsymbol{b}_1$ is \emph{intrinsically nonlinear}, since if both $\boldsymbol{b}_1$ and $\boldsymbol{b}_1'$ individually satisfy the constant-$|\boldsymbol{b}|$ constraint, their superposition $\boldsymbol{b}_1 + \boldsymbol{b}_1'$ generally violates that constraint, so linear superposition fails and challenges the conventional Fourier decomposition employed in Alfv\'enic turbulence studies. Second, the Alfv\'en speed is uniquely defined, because $\boldsymbol{b}_0$ is fixed by the constant far field and, expressed in Alfv\'en units, directly represents the unperturbed Alfv\'en velocity, eliminating the ambiguity inherent in classical ensemble-averaged definitions of the background field. Third, field-line twisting is non-trivial, because Alfv\'en's theorem dictates that magnetic field topology is preserved in ideal MHD and simultaneously maintaining $\delta|\boldsymbol{B}|\ll|\delta\boldsymbol{B}|$ with a constant unperturbed far field $\boldsymbol{B}_0$ necessitates non-trivial twisting to preserve quasi-constant field-line density, i.e. $|\boldsymbol{B}|$.

\subsection{Subtleties of the Alfv\'enon Model}\label{subsec:subtleties}

The construction of the Alfv\'enon model in Section~\ref{sec:model} involves operations in Fourier space, and thus inherits limitations associated with spectral methods, most notably the Gibbs phenomenon. The algorithm naturally generates sharp gradients at the edges of the perturbed region, which cannot be accurately represented due to spectral ringing. These overshoots violate the constant-$|\boldsymbol{B}|$ constraint and are subsequently suppressed by the iterative procedure, thereby shifting spectral energy toward high-$k$ modes. This effect also introduces complications in MHD simulations: the LAPS code advances in Fourier space and therefore cannot properly resolve strong discontinuities, while the dealiasing procedure suppresses high-$k$ components, inevitably affecting the evolution of the Alfv\'enon. For the parameters adopted in this study, such effects remain modest; however, for larger-amplitude Alfv\'enons, the accumulation of spectral energy at high wavenumbers becomes more pronounced, potentially leading to numerical instabilities.

As noted already in Section~\ref{sec:equations}, a more fundamental limitation concerns the constant-$|\boldsymbol{B}|$ constraint itself. We enforce constant $|\boldsymbol{B}|$ only as a \emph{soft} constraint because a strictly localized solution with exactly constant $|\boldsymbol{B}|$ is incompatible with magnetic-flux conservation in a finite domain. As argued in \citet{shi_analytic_2024} [see their Eq.~(10)], a cylindrical tube aligned with the $x$-axis and enclosing the perturbation must have the same total magnetic flux through every cross-section; however, any finite-amplitude reduction or reversal of $B_x$ inside the perturbed region removes axial flux that must be compensated elsewhere. The compensation appears as a slight compression or enhancement of neighboring field lines, and hence as a small departure from perfectly constant $|\boldsymbol{B}|$, unless strict localization is relaxed. The Alfv\'enon model is therefore most accurately regarded as a controlled quasi-constant-$|\boldsymbol{B}|$ and quasi-solitary realization of the ideal solitary Alfv\'en-wave formulation, consistent with the two \emph{soft} constraints introduced in Section~\ref{sec:model}.

\begin{figure}[htp!]
    \centering
    \includegraphics[width=1.0\columnwidth]{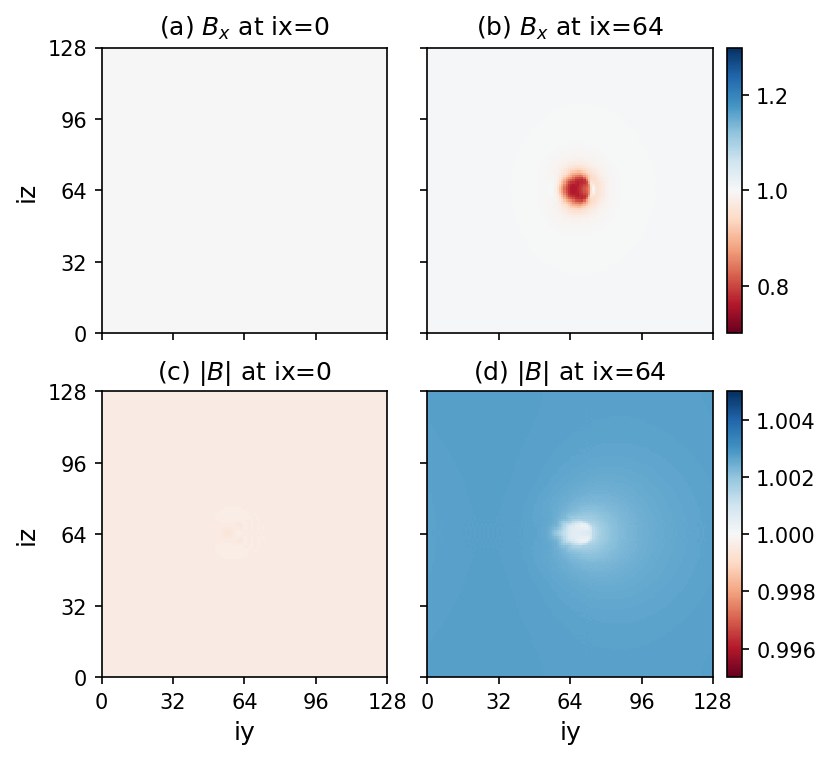}
    \caption{2D slices of $B_x$ and $|\boldsymbol{B}|$ at $ix = 0$ and $64$.}
    \label{fig:flux}
\end{figure}

Figure~\ref{fig:sim}(c1) is a manifestation of this effect in the $x$-$y$ plane. Since $V_A = |\boldsymbol{B}|/\sqrt{\rho}$ and $\rho=1$ in the initial condition, the panel effectively visualizes $|\boldsymbol{B}|$, which is slightly compressed (by $\sim 0.3\%$) in the slab containing the solution. Figure~\ref{fig:flux} illustrates this effect in greater detail through 2D slices of $B_x$ and $|\boldsymbol{B}|$ in both the unperturbed ($i_x=0$) and perturbed ($i_x=64$) regions. In the unperturbed region (panel a), $\boldsymbol{B}$ closely approximates $\boldsymbol{B}_0$ with only minor deviations ($\sim$ 0.1\%). In the perturbed region (panel b), the local reduction in $B_x$ is compensated by a slight enhancement of $B_x$ in the surrounding area, ensuring that the total magnetic flux is strictly identical across the two panels---a direct consequence of the strictly enforced solenoidal condition $\nabla\cdot\boldsymbol{B} = 0$. Panels (c) and (d) show the corresponding $|\boldsymbol{B}|$ distributions, revealing that the perturbed region slightly compresses the neighboring field lines and thereby increases the local field-line density, i.e.\ $|\boldsymbol{B}|$, by $\sim$ 0.3 \%. This compression is fully consistent with Fig.~\ref{fig:sim}(c1).

This effect is expected to diminish if the solution is constructed in a larger domain while preserving the size of the perturbed region. Solitary Alfv\'en wave packets with $B_x$ reversals are therefore inherently space-filling, which may have interesting consequences for the solar wind. Within the bounded coronal hole outflow, the presence of switchbacks should compress the surrounding field lines while preserving the total magnetic flux through each cross-section. Observations have consistently shown that $|\boldsymbol{B}|$ decreases more slowly than $R^{-2}$ in near-Sun coronal hole outflows \citep{bale_highly_2019,huang_solar_2024}, deviating from the spherical expansion trend expected from remote sensing. The space-filling nature of switchbacks may therefore directly account for this deviation. Future work will investigate this in detail.

\subsection{Conclusions}

To the best of our knowledge, the Alfv\'enon represents the first numerical realization of a solitary Alfv\'en wave packet. The Alfv\'enon exhibits nontrivial three-dimensional twisting of magnetic field lines while preserving quasi-constant $|\boldsymbol{B}|$. Its complex structure and coherent finite-time propagation suggest that fundamental aspects of Alfv\'en wave physics remain incompletely understood, more than eight decades after their original conceptualization. Moreover, the ubiquity of solitary Alfv\'en waves in the highly magnetized solar corona suggests that such localized structures may be an important form of Alfv\'enic fluctuations in astrophysical environments.

Despite its simplicity, the iterative algorithm yields nontrivial results. Two natural follow-up studies arise: (1) investigating the dependence of the Alfv\'enon structure on the initial amplitude $A$. The nontrivial field-line twisting revealed here suggests novel behaviors of magnetic fields in highly magnetized plasmas where $|\boldsymbol{B}|$ is constrained to remain constant. Our iterative algorithm thus provides a foundation for future investigations of solitary Alfv\'en wave physics across a variety of contexts. (2) Conducting direct MHD simulations of Alfv\'enon collisions. Nearly all MHD turbulence phenomenologies depend on the interaction of counter-propagating Alfv\'en wave packets \citep[see e.g.][]{kraichnan_inertialrange_1965,iroshnikov_turbulence_1964,goldreich_toward_1995,velli_turbulent_1989,boldyrev_spectrum_2005,boldyrev_spectrum_2006,chandran_parametric_2018,chandran_reflection-driven_2019}. Our model, for the first time, enables direct simulation of such collisions in isolated setups. Both topics will be addressed in forthcoming companion papers.

Finally, we note that spherically polarized Alfv\'en waves---and hence Alfv\'enons---are also exact solutions of relativistic MHD \citep{mallet_exact_2021}. In strongly magnetized environments where the Alfv\'en speed approaches the speed of light, relativistic Alfv\'enons could transport energy far more efficiently than classical shear Alfv\'en waves via their ultra-relativistic jets. Recent studies have proposed Alfv\'en waves as viable drivers of fast radio bursts (FRBs) from magnetars \citep{yuan_alfven_2021,yuan_magnetar_2022,kumar_frb_2020,long_fast_2025,chen_alfven_2025}. Exploring the behavior, stability, energy transport, and collision properties of relativistic Alfv\'enons represents an exciting direction for future research.

% \clearpage

% \vspace{2em}
% The python implementation of iterative algorithm is publicly available at \url{https://github.com/huangzesen/Helmholtz_algo_public}.

%% Please use the acknowledgment and contribution environments. This will 
%% be anonomyized when the "anonymous" style option is used. 
\begin{acknowledgments}
Z.H. thanks Benjamin Chandran, John. W. Belcher, Melvyn Goldstein, Margaret G. Kivelson, Krishan Khurana, Yingdong Jia, and Robert Strangeway for stimulating discussions. Claude AI and ChatGPT were used during the preparation of this work for exploratory brainstorming, language editing, and code-debugging assistance. In particular, the alternating-projection formulation of the numerical algorithm was developed during exploratory interactions with Claude AI; the authors subsequently formulated, implemented, tested, and interpreted the method, independently checked the AI-assisted material, and are fully responsible for the scientific content and conclusions of the manuscript. This work is supported by NASA HTMS 80NSSC20K1275 and NASA AIAH 80NSSC25K0386. C.S. acknowledges supported from NSF SHINE 2229566 and NASA ECIP 80NSSC23K1064.
\end{acknowledgments}

\begin{contribution}
Z.H. conceived the idea, designed the study, conducted the MHD simulations, and wrote the manuscript. M.V. contributed to the conceptual framework and provided critical guidance throughout the study. C.S. provided technical support for the simulation code and contributed to the conceptual development. Y.D. contributed to the theoretical framework. All authors reviewed and revised the manuscript.
\end{contribution}

% \clearpage

% \facilities{HST(STIS), Swift(XRT and UVOT), AAVSO, CTIO:1.3m, CTIO:1.5m, CXO}

% \software{}

\clearpage

\appendix

\section{3D Dependence of Simply Connected Magnetic Solitary Solution}\label{appendix:A}

Let $\mathbf{V}:\mathbb{R}^3 \to \mathbb{R}^3$ be a vector field that approaches a constant value at spatial infinity:
\[
\mathbf{V}(\mathbf{x}) \to \mathbf{V}_\infty 
\quad\text{as }\|\mathbf{x}\|\to\infty.
\]
No assumptions are imposed on the divergence, curl, or magnitude of $\mathbf{V}$.  
Suppose that in some coordinate system $(q_1,q_2,q_3)$ the field depends on only two variables:
\[
\mathbf{V}(q_1,q_2,q_3)=\mathbf{V}(q_1,q_2).
\]

\subsection*{Claim}
A nontrivial isolated configuration of this type is only possible if the suppressed coordinate  
$q_3$ parametrizes a \emph{compact loop} in physical space.  
If $q_3$ is unbounded, the only vector field satisfying the asymptotic condition is the trivial uniform field $\mathbf{V}_\infty$.

\subsection*{Argument}

For each fixed $(q_1,q_2)$, the field remains constant as $q_3$ varies.  
Define the set
\[
\Gamma_{(q_1,q_2)} = \{ (q_1,q_2,q_3) : q_3 \in I \},
\]
where $I$ is the range of $q_3$.  
All points in $\Gamma_{(q_1,q_2)}$ share the same field value $\mathbf{V}(q_1,q_2)$.

\medskip

\textbf{Case 1: $q_3$ unbounded.}  
If $q_3$ ranges over an unbounded interval, then $\Gamma_{(q_1,q_2)}$ contains points with arbitrarily large Euclidean norm.  
Along this direction,
\[
\mathbf{V}(q_1,q_2,q_3) = \mathbf{V}(q_1,q_2)
\quad\text{for all } q_3.
\]
If $\mathbf{V}(q_1,q_2) \neq \mathbf{V}_\infty$, then the field fails to approach $\mathbf{V}_\infty$ along these unbounded lines, contradicting the assumed asymptotic behavior.  
Thus the condition at infinity forces
\[
\mathbf{V}(q_1,q_2)=\mathbf{V}_\infty
\quad\forall (q_1,q_2),
\]
so $\mathbf{V}$ is constant everywhere.

\medskip

\textbf{Case 2: $q_3$ compact.}  
If $q_3$ parametrizes a topologically closed region (e.g.\ an angular coordinate), then each set $\Gamma_{(q_1,q_2)}$ is bounded.  
Nontrivial dependence on $(q_1,q_2)$ does not produce directions along which the field propagates to infinity while remaining fixed.  
Therefore the field may coincide with $\mathbf{V}_\infty$ outside a sufficiently large ball while retaining nontrivial structure in a compact region.

\subsection*{Conclusion}

A vector field on $\mathbb{R}^3$ that becomes uniform at infinity and depends on only two coordinates can exhibit nontrivial localized structure only if the remaining coordinate is spatially bounded, forming topologically closed regions.  
If the unused coordinate is unbounded, the asymptotic condition forces the field to be identically equal to the far-field constant $\mathbf{V}_\infty$.

\section{Field Line Tracer}\label{appendix:B}

A Runge-Kutta method is employed to trace the field lines. Given the magnetic grid $\{B_x,B_y,B_z\}$ on a periodic box of size $\mathbf{L}$, each component is trilinearly interpolated to an arbitrary point $\mathbf{x}$:
\[
\mathbf{B}(\mathbf{x}) = \bigl(B_x(\mathbf{x}),\,B_y(\mathbf{x}),\,B_z(\mathbf{x})\bigr),
\]
with indices wrapped periodically before interpolation. The local direction field is then
\[
\hat{\mathbf{b}}(\mathbf{x}) = 
\begin{cases}
\dfrac{\mathbf{B}(\mathbf{x})}{\lVert \mathbf{B}(\mathbf{x}) \rVert}, & \lVert \mathbf{B}(\mathbf{x}) \rVert > 10^{-12},\\[0.75em]
\mathbf{0}, & \text{otherwise}.
\end{cases}
\]

Starting from a seed $\mathbf{x}_0$, Runge–Kutta 4 (step size $h$) advances along this interpolated, unit-magnitude field:
\[
\begin{aligned}
\mathbf{k}_1 &= \hat{\mathbf{b}}(\mathbf{x}_n),\\
\mathbf{k}_2 &= \hat{\mathbf{b}}\!\left(\mathbf{x}_n + \tfrac{1}{2} h \mathbf{k}_1\right),\\
\mathbf{k}_3 &= \hat{\mathbf{b}}\!\left(\mathbf{x}_n + \tfrac{1}{2} h \mathbf{k}_2\right),\\
\mathbf{k}_4 &= \hat{\mathbf{b}}(\mathbf{x}_n + h \mathbf{k}_3),\\
\mathbf{x}_{n+1} &= \operatorname{mod}\!\left(\mathbf{x}_n + \tfrac{h}{6}(\mathbf{k}_1 + 2\mathbf{k}_2 + 2\mathbf{k}_3 + \mathbf{k}_4),\, \mathbf{L}\right).
\end{aligned}
\]

Each step samples the trilinearly interpolated $\hat{\mathbf{b}}$ at the intermediate RK4 positions and wraps the result back into the periodic domain, producing the traced field line forward and/or backward from the seed.

\section{Dealiasing in LAPS} \label{appendix:C}

LAPS provides two dealiasing options, summarized below.

Option 1 (circular-padding / $2/3$ rule) applies a sharp spectral cutoff, retaining only modes with $k \le \tfrac{2}{3}k_{\max}$:
\[
G_1(k)=
\begin{cases}
1, & k \le \tfrac{2}{3}k_{\max},\\
0, & k > \tfrac{2}{3}k_{\max}.
\end{cases}
\]

Option 2 (smoothing filter) applies a smooth rational filter with $\theta=\pi k/k_{\max}$:
\[
G_2(\theta;\alpha)=\frac{a_j+b_j\cos\theta+c_j\cos(2\theta)}{1+2\alpha\cos\theta},
\quad
a_j=\frac{5+6\alpha}{8},\;
b_j=\frac{1+2\alpha}{2},\;
c_j=-\frac{1-2\alpha}{8}.
\]
As $\alpha \to 0.5$, the filter becomes less dissipative (closer to unity), so high-$k$ suppression weakens.

\begin{figure}[!htp]
    \centering  
    \includegraphics[width=0.5\textwidth]{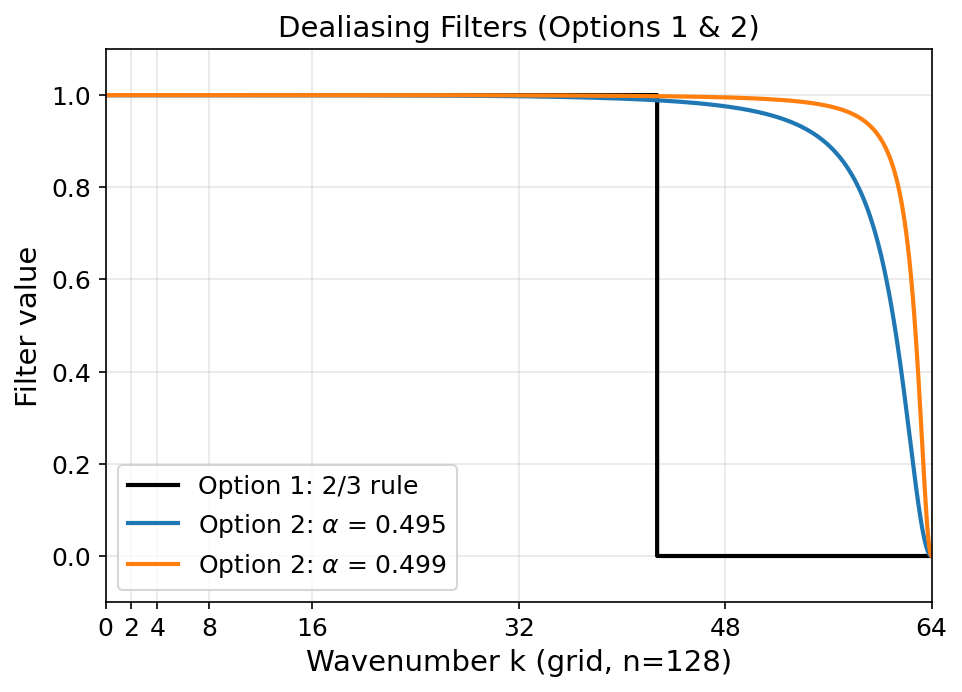}
    \caption{Dealiasing filters for a $128^3$ grid ($k_{\max}=64$). Option~1 applies a sharp $2/3$ cutoff at $k=2k_{\max}/3$, while Option~2 uses the smooth rational filter with $\alpha=0.495$ and $\alpha=0.499$, producing a gradual high-$k$ rolloff.}
\end{figure}

\clearpage

\bibliography{references}{}
\bibliographystyle{aasjournalv7}

\end{document}